# A Systemic Pathological Network Model and Combinatorial Intervention Strategies for Alzheimer's Disease

Author:She Xutong

Affiliation:Nanjing University of Science and Technology

Corresponding Author: She Xutong

Email: shexutong@njust.edu.cn

Abstract

Alzheimer's disease (AD) persists as a paramount challenge in neurological research, characterized by the pathological hallmarks of amyloid-β (Aβ) plaques and neurofibrillary tangles composed of hyperphosphorylated tau. This review synthesizes the evolving understanding of AD pathogenesis, moving beyond the linear amyloid cascade hypothesis to conceptualize the disease as a cross-talk of intricately interacting pathologies, encompassing Aβ, tau, and neuroinflammation as the foundation of phase-adapted pathological network model. This evolving pathophysiological understanding parallels a transformation in diagnostic paradigms, where biomarker-based strategies—such as the AT(N) framework—enable early disease detection during preclinical or prodromal stages. Within this new landscape, while anti-Aβ monoclonal antibodies (e.g., lecanemab, donanemab) represent a breakthrough as the first disease-modifying therapies, their modest efficacy underscores the limitation of single-target approaches. Therefore, I explore the compelling rationale for combination therapies that simultaneously target Aβ pathology, aberrant tau, and neuroinflammation. Looking forward, I emphasize emerging technological platforms—such as gene editing and biophysical neuromodulation—in advancing precision medicine. Ultimately, the integration of early biomarker detection, multi-target therapeutic strategies, and AI-driven patient stratification charts a promising roadmap toward fundamentally altering the trajectory of AD. The future of AD management will be defined by preemptive, biomarker-guided, and personalized combination interventions.

# 1.Introduction

Alzheimer's disease (AD) stands as the predominant form of dementia, presenting significant and escalating global challenges. It is characterized by various pathological markers in the brain——large numbers of amyloid plaques surrounded by neurons containing neurofibrillary tangles, vascular damage from extensive plaque deposition, and neuron cell loss.[1] According to the World Alzheimer Report 2015, global dementia-related costs were estimated at US$606.7 billion in 2010 and US$817.9 billion in 2015 (adjusted for prevalence and World Bank country classifications).[2] Advancing research and improving treatment and care pathways for AD are therefore critical to mitigating societal and familial burdens.

The amyloid cascade hypothesis has dominated research and drug development efforts for decades, amassing impressive amounts of supporting evidence.[3] Despite significant progress in understanding the biological generation and clearance of amyloid beta (Aβ), this paradigm is found wanting by the fact that 33 phase 3 clinical trials of drugs targeting Aβ have failed to slow cognitive decline in AD,[4] coupled with a discrepancy between extent of plaque pathology and neurodegeneration.[5,6] This implies us that the pathogenesis of AD is far more intricate than the malfunction of a single molecular pathway. Current antibody-targeted Aβ pathway fall to capture the aetiological complexity and clinical heterogeneity of patients, underscoring the necessity for a network-based and anti-reductionist disease view.[7,8] Data suggests that evolving multi-factor disease models will better underpin the search for more effective strategies to treat the disease.[8] Therefore, I establish a systematic model, positing that the pathogenesis of AD is not a linear cascade but a destructive cycle composed of three interconnected subsystems. Building on this framework, I propose diagnostic strategies that transcend conventional Alzheimer's theories and align with systems biology principles. The paper ultimately presents a novel therapeutic paradigm characterized by "multi-target collaborative intervention" and "early precision intervention". Through synthesizing and reinterpreting existing evidence, I aim to provide a theoretical foundation and strategic direction for developing more effective and resilient AD prevention and treatment frameworks.

## 2. Mechanisms

The pronounced heterogeneity of AD arises from multidimensional variation across etiologic drivers, clinical phenotypes, and neuropathologic profiles.[9] As a result, constructing an integrative framework that seamlessly links genetic underpinnings, molecular mechanisms, and clinical expression remains formidable. Current methodological and evidentiary limitations further constrain a comprehensive understanding of AD pathophysiology.[10] This section moves beyond the traditional, linear enumeration of hypotheses to construct a multi-level, interactive pathological network model of Alzheimer's Disease (AD). When constructing the systemic pathological network of Alzheimer's disease (AD), two core principles must be coordinated: systemic interconnectedness and hierarchical organization. Systemic interconnectedness emphasizes the indispensable and interdependent nature of all pathological components, while hierarchical organization reveals the differential influence and temporal sequence of actions among components within the network. Building on this foundation, this study synthesizes existing evidence to propose a AD pathological network model: amyloid-β (Aβ), tau, and neuroinflammation. Within this network, the synergy between Aβ and tau acts as the primary disease driver, while neuroinflammation serves as a critical amplifier. This model aims to clearly delineate the primary and secondary roles of each subsystem in synergy, providing a framework for subsequent diagnose and treatments.

## 2.1 Interplay Between Aβ and Tau

### 2.1.1 Amyloid Beta drives Tau

Several hypotheses related to the pathogenesis of AD were studied,[11] and the amyloid-β (Aβ) cascade and the hyperphosphorylation of tau protein are the two main hypotheses. The conventional view regards Aβ as the upstream trigger, while Tau-mediated neurofibrillary tangles are considered the downstream pathway. The Aβ cascade hypothesis states that Aβ, deposited in the form of neuroinflammatory plaques, induces AD by damaging neuronal cells.[12] Aβ facilitates the development of AD and initiates a deleterious cascade involving tau pathology and neurodegeneration.[13] Nonetheless, more than 200 drug candidates targeting various key aspects of the amyloid-cascade model almost uniformly failed to provide benefits in clinical trials, with some leading to worsening of symptoms and/or other adverse outcomes.[14-17] These AD drugs examined in clinical trials and laboratory studies focused on the reduction of amyloid deposition and the clearance of Aβ oligomers and have not yielded satisfactory effects,[18,19] indicating that amyloid-β (Aβ) may not be the sole pathogenic protein in the pathogenesis of Alzheimer's disease (AD). The tau hypothesis posits that aberrant hyperphosphorylation, mislocalization, and aggregation of the microtubule-associated protein tau into neurofibrillary tangles (NFTs) precipitate collapse of the neuronal microtubule network, impair axonal transport, disrupt synaptic function, and ultimately drive neuronal death[20,21] — constituting principal mechanisms underlying neurodegeneration and cognitive decline in Alzheimer's disease (AD).[21] Intraneuronal tangles containing hyperphosphorylated tau are a hallmark of AD pathology,[22] and tau is a mediator of Aβ cytotoxicity.[19] Therefore, the pathological changes to tau in AD also received attention. Nonetheless, extensive studies focused solely on the neurotoxicity of Aβ or tau have not shown significant efficacy in the treatment of AD.[23,24] Therefore, focusing solely on the role of Aβ or tau in AD lesions while ignoring the interaction between Aβ and tau may not be entirely correct. Clinicopathological observations reveal discrepancy between plaque burden and cognition:[25] some older adults exhibit abundant Aβ plaques at autopsy despite preserved premortem cognitive function,[26] whereas others with demonstrable cognitive impairment may show relatively modest plaque burden.[27,28] These findings suggest that Aβ deposition is likely not sufficient for clinical expression, to some extant challenging the original Amyloid Cascade Hypothesis as upstream position.

Studies proposes that symptomatic disease is produced obligatorily by convergence with additional factors—such as tau pathology, neuroinflammation, vascular injury, and variability in cognitive reserve.[29] Emerging evidence underscores their synergistic cooperation, wherein amyloid accumulation may facilitate the aggregation[30] and broader dissemination of tau,[31,32] and toxic state of tau can enhance Aβ toxicity via a feedback loop.[33,34] ) It is aligned with nonlinear acceleration of tau pathophysiology throughout the natural course of Alzheimer's disease.[35,36] Neuroimaging and network analyses have revealed that there are two pivotal Aβ-tau interactions in the natural history of AD. The first occurs when neocortical Aβ emerges within multiple neocortical and limbic regions connected to the entorhinal cortices. This remote, connectivity-mediated interaction between Aβ and EC tau may induce tau to undergo biophysical changes that propel it to spread out of the entorhinal areas and into nearby, connected regions in the hippocampus, amygdala, and basal temporal cortices.[37] In the subtemporal gyrus where Aβ and tau coexist, Aβ plaques induce localized neuroinflammation by activating microglia. This inflammatory environment renders neurons vulnerable to tau protein damage. The second pivotal moment occurs when tau neurofibrillary changes reach the ITG, where tau can locally interact, for the first time, with pre-existing Aβ, catalyzing widespread tau propagation into Aβ-positive and ITG-connected neocortical regions whose degeneration ultimately gives rise to dementia. Mediation analysis confirmed that the Aβ-induced increases in functional connectivity accounts for 5% to 25% of the variance in tau accumulation effects.[38]

Aβ and tau may interact via intermediate materials, such as kinases (e.g., GSK-3β, CDK-5 and ERK),[39]. The removal of Aβ or tau alone does not completely terminate the interaction pathway, which continues to play a role in the acceleration of the pathological process ,[40,41] in which these proteins amplify toxic effects rather than a strictly hierarchical mode of interaction.[41] Specifically, Aβ orchestrates tau hyperphosphorylation primarily by activating key cellular kinases.[42] Soluble Aβ oligomers engage neuronal surface receptors such as NMDA and insulin receptors, triggering intracellular signaling cascades that lead to calcium influx and oxidative stress. These events activate stress-responsive kinases, including glycogen synthase kinase-3β (GSK-3β) and cyclin-dependent kinase 5 (CDK5).[43] GSK-3β, in particular, is a major tau kinase whose activity is upregulated by Aβ-induced disruption of calcium homeostasis and mitochondrial function.[44] Once activated, GSK-3β phosphorylates tau at multiple epitopes (e.g., Ser396,Thr231), reducing its affinity for microtubules and promoting its dissociation.[45] The unique environment Aβ plaques created facilitates tau misfolding,[46] oligomerization,[47] and eventual aggregation into neurofibrillary tangles (NFTs).[48,49] That is to say, Pre-aggregated Aβ acts as a driver, significantly enhancing the fibrillarization and aggregation capacity of tau protein, providing direct biochemical evidence supporting the 'prion-like' transmission of Aβ in promoting Tau pathology.[30]

In summary, when tau is activated by Aβ, it forms dense connections with the temporal lobe's inferior gyrus and extensive brain regions involved in memory, visual processing, and semantic cognition, thereby accelerating and expanding its diffusion across neural networks.[37] This framework delineates a critical period in AD pathogenesis, which is initiated by the emergence of Aβ in EC-connected regions. This period continues with the subsequent spread of tau from the EC to medial temporal and limbic areas, and culminates in the interaction between Aβ and tau within the ITG hubs. The well-connected nature of these ITG hubs then facilitates the widespread neocortical propagation of tau. The efficacy of Aβ-lowering drugs may depend on the timing of their delivery[50], and the critical period defined here can now be evaluated as a potential therapeutic window. Generally, Aβ and Tau in some extent have synergy effects. Therefore, suppression of the interaction may be of more practical significance than simply focusing on the neurotoxicity of Aβ or tau alone.[51]

*2.1.2 Tau Drives Amyloid Beta*

As outlined in Section 2.1.1, Aβ plays a pivotal role in the early stages of Alzheimer's disease (AD) by facilitating the outward diffusion of tau protein from the entorhinal cortex and creating a "toxic environment" conducive to tau pathology. However, tau is not merely a passive downstream executor. Growing evidence indicates that pathological changes in tau protein actively feed back to exacerbate Aβ accumulation and toxicity, collectively driving AD progression.

The toxic effects of Aβ originate from its effective binding to NMDARs.[52] Activation of extrasynaptic NMDARs induces neuronal death, whereas synaptic NMDAR activation promotes cell survival. Aβ oligomers bind to and activate toxic pathways by altering the distribution of NMDAR subunits,[53-55] in line with glutamate hypothesis. Nonetheless, in the absence of tau, Aβ cannot induce significant neurotoxicity[56] or effectively activate downstream pathological kinases such as GSK3β and p38γ.[57] Animal model studies have provided crucial evidence for tau's decisive role in Aβ toxicity. In transgenic mice with Alzheimer's disease (AD), reducing tau protein levels prevents cognitive impairment caused by Aβ overexpression; conversely, reintroducing tau protein restores neurons' sensitivity to Aβ toxicity, thereby rendering neurons more susceptible to the deleterious effects of Aβ. Additionally, excessive expression of Fyn kinase (whose function depends on tau presence) exacerbates Aβ-induced cognitive damage.[58] Collectively, these findings establish that hyperphosphorylated tau exacerbates the toxic effects of Aβ by activating specific kinases.

More importantly, hyperphosphorylated tau further increases Aβ production and accumulation by inhibiting brain Aβ clearance pathways (e.g., downregulating insulin-degrading enzyme IDE expression) and promoting abnormal cleavage of amyloid precursor protein (APP). Mislocalized and hyperphosphorylated tau impairs axonal transport,[59] including the trafficking of amyloid precursor protein (APP) and its processing enzymes.[60] This disruption favors amyloidogenic APP cleavage by BACE1 and γ-secretase, increasing Aβ production.[61] In addition, tau aggregates contribute to synaptic failure and neuronal hyperexcitability, which elevate neuronal activity and promote Aβ release. Notably, extracellular tau species—released from degenerating neurons—can act as seeds that promote Aβ aggregation by cross-seeding or by stabilizing toxic Aβ oligomers.[62] This process stands as a interconnected destructive cycle to drive the progression of AD, leaving both antagonistic and synergistic interactions between Aβ and tau. When they coexisting, tau blocks Aβ-mediated hyperexcitability, making neuronal silencing the dominant phenotype and leading to severe functional inhibition of neural circuits. But when they stand separately, isolated Aβ may induce neuronal hyperexcitability, standalone tau can suppress neuronal activity silencing.[63] At that time, tau suppression was incapable of rescuing tau-dependent neuronal silencing in two APPxtau models.[64] The core of this antagonistic effect lies in tau: without tau, the toxicity of Aβ manifests solely as hyperactivity, failing to induce the characteristic circuit dysfunction observed in Alzheimer's disease (AD).

Clinical imaging and experimental evidence jointly support the synergistic interaction between Aβ and tau. In the brains of Alzheimer's disease (AD) patients, Aβ accelerates the diffusion pattern of tau pathology, which typically follows Braak staging.[65] When Aβ and tau coexist, they synergistically cause reduced brain metabolism, regional atrophy, and significantly accelerated cognitive decline.[66,67] In animal models, crossbreeding Aβ transgenic mice with tau transgenic mice leads to earlier tau tangles, neuronal loss, and more severe cognitive impairment compared to single-transgenic mice.[68] Injecting Aβ into tau model mice accelerates tau aggregation and diffusion to synaptic regions,[69] while injecting tau seeds into Aβ model mice enhances pathological tau accumulation around plaques.[70] Furthermore, novel mouse models have demonstrated that reactive astrocyte proliferation induced by toxic molecules such as Aβ and p-Tau significantly enhances the accumulation of Aβ plaques and neurofibrillary tangles (NFTs), thereby exacerbating Alzheimer's disease (AD) pathology including neuronal loss and memory impairment.[71]

In summary, tau protein not only mediates Aβ neurotoxicity but also actively exacerbates Aβ pathology through multiple mechanisms: disrupting APP transport, promoting Aβ formation, and acting as a "seed" to facilitate Aβ aggregation. This forms a tight resonance and closed loop with the "Aβ-driven tau" mechanism described in Section 2.1.1. These two processes are not merely upstream-downstream relationships but synergistic association between Aβ and tau in the progression of Alzheimer's disease.[67] Several lines of experimental and clinical evidence now indicate that the role of Aβ is more complex in that its presence enhances tau phenotypes throughout the disease course.[72] It implies that targeting a single protein (Aβ or tau) alone may be insufficient for effective disease control in AD treatment, suggesting that combined intervention strategies or therapies targeting their interaction nodes may hold greater therapeutic potential.

## 2.2 Microglia in Neuroinflammation

### 2.2.1 Neuroinflammation

Neuroinflammation, characterized by chronic activation of microglia and astrocytes alongside pro-inflammatory mediator release, represents a sustained immune response within the

Central Nervous System(CNS).[73,74] Microglia, the brain's resident immune cells, play a dual role. Under physiological conditions, they support synaptic homeostasis.[75] In AD, they are activated by Aβ[76] and attempt to clear plaques via receptors like TREM2 (Triggering Receptor Expressed on Myeloid cells 2). TREM2 is a key receptor on the surface of microglia. In the AD brain, it senses "danger signals" such as Aβ plaques and neuronal debris and binds to ligands like ApoE and Aβ. However, chronic activation leads to phagocytic dysfunction, sustained release of pro-inflammatory cytokines (e.g., IL-1β, TNF-α),[76] and NLRP3 inflammasome activation,[77] which paradoxically promotes further Aβ aggregation.[78] Furthermore, these inflammatory mediators directly exacerbate tau pathology; for instance, TNF-α can activate GSK-3β,[79] a kinase that drives tau hyperphosphorylation.[78 80] This creates a self-reinforcing cycle where Aβ and tau jointly sustain a neurotoxic environment, and inflammation, in turn, facilitates tau propagation and synaptic damage. Excessive activation of the complement cascade may also drive aberrant microglia-mediated synaptic pruning, further aggravating neurodegeneration.[73] Given the complex role of TREM2 protein and its potential as a promising target for the prevention of Alzheimer's disease (AD), it is essential to investigate how TREM2 mediates the activation, migration, and clearance of Aβ.

Upon ligand of TREM2 binding, the ITAM motif of its partner DNAX-activation protein 12 (DAP12) becomes phosphorylated, leading to the recruitment and activation of the cytoplasmic kinase Syk. This, in turn, activates the PI3K-Akt pathway, the PLCγ pathway, and the Vav-Rac pathway. The PI3K-Akt pathway promotes cell survival and proliferation. The PLCγ pathway generates IP3 and DAG, inducing intracellular calcium release, which drives cytoskeletal reorganization and provides the mechanical force for cell migration. The Vav-Rac pathway directly activates the small GTPase Rac, which orchestrates actin polymerization and the formation of lamellipodia, providing the structural basis for cell migration and phagocytosis. Subsequently, microglia migrate directionally along the Aβ concentration gradient toward the plaques. They extend lamellipodia to engulf Aβ, forming phagosomes. These phagosomes then fuse with lysosomes, leading to the degradation of Aβ. Furthermore, TREM2 signaling can suppress excessive activation of the NF-κB pathway via mechanisms involving PI3K-Akt, thereby preventing harmful neuroinflammation. The core mechanism is illustrated in Figure 1. If TREM2 function is impaired (e.g., due to genetic variants), microglia fail to be properly activated, migrate, and clear Aβ, resulting in a loss of their neuroprotective function and ultimately exacerbating AD pathology.

Despite substantial evidence linking inflammation to Alzheimer's disease (AD) pathology, limited longitudinal data and the paucity of very-early biomarkers hinder precise delineation of causal directionality and temporal onset.[81] One model posits that abnormal aggregation of Aβ oligomers or tau initially activates glial cells, precipitating an inflammatory response. An alternative "inflammation-first" hypothesis proposes that early dysregulation of innate immunity precedes and promotes the emergence and spread of Aβ and tau pathologies.[74] These two perspectives may both reflect a complex bidirectional interplay between inflammation and core pathologies throughout the protracted disease course.[82] A pivotal shift is now underway, with recent investigations concentrating on the collapse of communication mechanisms between brain cells. It is proposed that the breakdown in neuron-glia crosstalk, especially involving astrocytes and microglia, may underpin disease advancement.[83] Proteomic studies have identified novel and major facilitator such as AHNAK(Adductin-H-N-A-K), underscoring a paradigm shift in AD research. This research marks a move beyond a narrow focus on Aβ and tau toward a "systems biology" perspective that seeks to deconstruct the entire dysregulated network of the disease. Scientists are no longer satisfied with hunting for a single pathology but are now committed to elucidating the intricate interplay of pathologies. [83] There is also research shows that, the implementation of strategies targeting CD44 and CD33, enhancing the expression of HLA-DR, P2RY12, and ApoE, and modulating microenvironmental signals may contribute to slowing or reversing the pathological progression of Alzheimer's disease, thereby providing a scientific rationale for developing novel therapeutics.[84]

Given its complex dual roles, targeting neuroinflammation presents a unique therapeutic challenge: how to suppress its chronic, destructive outputs while preserving its acute, protective functions. This complexity explains the failure of broad anti-inflammatory drugs (e.g., NSAIDs) in AD trials. The emerging goal is not wholesale suppression, but precision immunomodulation. Strategies now aim to reprogram microglial states—for instance, by enhancing protective pathways like TREM2 signaling to boost Aβ clearance, or by targeting specific inflammatory nodes (e.g., NLRP3, CD33) to curb cytokine production without compromising phagocytosis. The functional state of microglia is thus a critical determinant of therapeutic efficacy, influencing even the response to anti-Aβ immunotherapies.

This nuanced view is part of a broader paradigm shift towards a "systems biology" perspective of AD. Research is moving beyond a narrow focus on individual pathologies to deconstruct the entire dysregulated cellular network. This includes understanding how metabolic disturbances (e.g., in obesity or diabetes) can predispose microglia to a dysfunctional, pro-inflammatory state, thereby lowering the brain's resilience to Aβ and tau. This systems-level understanding underscores that successful therapeutic intervention will require a multi-target approach. It necessitates combining Aβ- and tau-directed therapies with immunomodulatory agents that are precisely timed and tailored to an individual's neuroinflammatory status.

In conclusion, neuroinflammation is a central pathogenic amplifier in AD, inextricably linking Aβ and tau pathologies into a coordinated destructive network. Its presence underscores the limitation of any single-pathway intervention. Taming this maladaptive immune response requires early, biomarker-guided detection of neuroinflammatory states and combination regimens that include precision immunomodulation, alongside amyloid and tau therapeutics, to effectively disrupt the disease cycle.

**2.2.2 Interplay among Aβ, Tau and Neuroinflammation**

As outlined in the previous section, neuroinflammation is intricately linked with both Aβ and tau pathologies. Building upon that foundational understanding, this section will delve deeper into the mechanistic interaction of pathways of Aβ, tau and neuroinflammation.

While neuroinflammation initially represents a protective response to central nervous system (CNS) insults, its chronic activation in Alzheimer's disease (AD) profoundly exacerbates core pathological processes, establishing a self-perpetuating cycle of neurodegeneration.[73] This maladaptive immune response not only fails to resolve underlying pathologies but actively amplifies them through multiple interconnected mechanisms. In this process, Aβ and tau act as principal pathological drivers that potently induce the activation of astrocytes and microglia,[42] which in turn release a spectrum of pro-inflammatory cytokines such as tumor necrosis factor alpha (TNF-α) and interleukin-1β (IL-1β), along with reactive oxygen species (ROS) and reactive nitrogen species (RNS),[85] thereby triggering neuroinflammation.[86] However, this inflammatory response has a dual role: while it may provide protective effects by enhancing Aβ degradation and clearance,[87] it can also lead to excessive production of Aβ and tau proteins,[88,89] ultimately inducing neurodegeneration and synaptic loss.

In addition to activating astrocytes and microglia, Aβ and tau proteins also exacerbate each other's pathological effects, thereby intensifying neuroinflammation. Pathological tau protein disrupts Aβ clearance mechanisms by at least two ways, leading to persistent Aβ accumulation. One way is that pathological tau protein activates microglia and astrocytes, sustaining a chronic inflammatory state and impairing microglia's capacity to engulf and clear Aβ.[90] Aβ progressively accumulates, and this accumulated Aβ sustains inflammatory responses by activating microglial NLRP3 inflammasomes and other pathways. The other way is that tau-laden neurons release factors that suppress microglial phagocytosis and reduce expression of Aβ-degrading enzymes such as neprilysin.[91,92] Furthermore, it has been proposed that extracellular tau species might promote Aβ aggregation through mechanisms such as cross-seeding or by stabilizing toxic Aβ oligomers.[90,93] The newly formed, structurally more stable Aβ aggregates (particularly oligomers and plaques) serve as potent activation signals for glial cells.[94] This enhanced stimulation promotes excessive activation of microglia and astrocytes, triggering inflammatory responses that release excessive pro-inflammatory cytokines and reactive oxygen species, thereby significantly exacerbating overall neuroinflammation.[95]

Neuroinflammation directly accelerates tau pathology by activating a network of kinases that drive tau hyperphosphorylation. Pro-inflammatory cytokines, particularly TNF-α and IL-1β, stimulate signaling pathways that activate several tau kinases, most notably glycogen synthase kinase-3β (GSK-3β)[96,97] and p38 mitogen-activated protein kinase (p38 MAPK).[98] For instance, TNF-α can activate GSK-3β via the PKCδ pathway, leading to phosphorylation of tau at pathological epitopes such as AT8 (Ser202/Thr205)[99] and PHF-1 (Ser396/404).[100] Moreover, inflammasome-derived IL-1β has been shown to enhance tau phosphorylation both in vitro and in vivo,[79,101] contributing to the formation of neurofibrillary tangles (NFTs). Critically, during the pathological progression of AD, Aβ and tau proteins may trigger neuroinflammation, which in turn acts as a partial driver of tau pathology, collectively forming an intricate, destructive pathological cycle. Conversely, tau pathology indirectly promotes neuroinflammation by impeding Aβ clearance,[102] creating another positive feedback loop. Pathological tau protein disrupts Aβ clearance mechanisms by at least two ways, leading to persistent Aβ accumulation. One way is that pathological tau protein activates microglia and astrocytes, sustaining a chronic inflammatory state and impairing microglia's capacity to engulf and clear Aβ.[90] This directly results in the persistent accumulation of Aβ, which subsequently drives and escalates inflammatory responses through mechanisms such as activating NLRP3 inflammasomes in microglia. The other way is that tau-laden neurons release factors that suppress microglial phagocytosis and reduce expression of Aβ-degrading enzymes such as neprilysin.[91,92] Furthermore, it has been proposed that extracellular tau species might promote Aβ aggregation through mechanisms such as cross-seeding or by stabilizing toxic Aβ oligomers.[90,93] The newly formed, structurally more stable Aβ aggregates (particularly oligomers and plaques) serve as potent activation signals for glial cells.[94] This enhanced stimulation promotes excessive activation of microglia and astrocytes, triggering inflammatory responses that release excessive pro-inflammatory cytokines and reactive oxygen species, thereby significantly exacerbating overall neuroinflammation.[95] At this point, Aβ, tau, and neuroinflammation form an inescapable destructive cycle, culminating in irreversible structural damage to brain networks, such as the abnormal clearance of synapses primarily mediated by the complement system. In the AD brain, chronic inflammatory conditions upregulate complement components such as C1q and C3.[103] These proteins tag synapses for elimination by engaging complement receptors (e.g., CR3) on microglia,[104] triggering excessive, activity-independent phagocytosis of synaptic structures[105]—a process analogous to developmental synaptic pruning but pathologically activated in the adult brain.[106] This complement-mediated synaptic stripping is further amplified by Aβ oligomers,[107] which can directly activate the classical complement pathway.[108] The resultant loss of excitatory synapses, particularly in the hippocampus and cortex,[109] is a major structural correlate of cognitive decline in AD. Thus, neuroinflammation transforms a physiological pruning mechanism into a destructive process that directly undermines neural circuit integrity and cognitive function.

In summary, neuroinflammation acts as a central pathological amplifier in AD, impairing Aβ clearance, promoting tau hyperphosphorylation, and driving complement-mediated synaptic loss. These mechanisms collectively underscore that inflammation is not a passive bystander but an active driver of disease progression. Targeting specific inflammatory pathways—while preserving beneficial immune functions—thus represents a critical therapeutic strategy for disrupting this destructive cascade.

## 2.3 An Integrated Network Model: From Linear Hypotheses to a Pathological Cascade

The preceding sections have detailed the principal pathological pathways in Alzheimer's disease (AD) as distinct entities. However, the repeated failure of therapies targeting single pathways underscores a fundamental insight: AD would rather be a syndrome driven by a dynamic, self-reinforcing pathological network than a disorder of a linear cascade.[82,110] This network view explains why the disease, once initiated, gains momentum and becomes progressively independent of its initial trigger. Converging proposal posit that neuroinflammation interplays with Aβ and tau synergistic effects, seeming to be a pivotal factor in AD's pathophysiology.[63,111]

The disease cascade is often catalyzed by an initial disruption in Aβ homeostasis, primarily impaired clearance in sporadic AD. The resulting accumulation of Aβ assemblies, particularly soluble oligomers, acts as a chronic "danger signal." As detailed in Section 3.1.1, these oligomers bind neuronal receptors, triggering intracellular calcium dysregulation and oxidative stress. This compromised neuronal environment is the first critical step in creating a permissive milieu for the amplification of downstream pathologies. It is important to note that at this stage, significant cognitive symptoms may be absent, representing the long preclinical phase of AD.[112]

The compromised neuronal environment and the presence of Aβ plaques potently activate microglia, primarily via receptors like TREM2 (Section 3.2.1). Initially, this is a protective response aimed at clearing debris. However, sustained activation leads to a maladaptive state characterized by two key failures: a diminished capacity for Aβ phagocytosis and an exaggerated release of pro-inflammatory cytokines (e.g., IL-1β, TNF-α).[113] This inflammatory milieu, as explored in Section 3.2.2, directly fuels the tau pathology cascade. For instance, TNF-α signaling can activate kinases like GSK-3β and p38 MAPK, which hyperphosphorylate tau, facilitating its dissociation from microtubules and misfolding.

This inflammation-driven hyperphosphorylation marks the critical transition of tau from a passive bystander to an active executing agent. As described in Section 3.1.2, pathological tau gains prion-like properties, enabling its trans-synaptic spread throughout brain networks.[114] This spread of tau pathology (Braak staging) correlates strongly with the emergence and progression of clinical symptoms. Furthermore, tau itself becomes a novel inflammatory stimulus, further activating glial cells and solidifying the chronic inflammatory state. This creates a destructive cycle, wherein tau-driven inflammation further perturbs neuronal and glial homeostasis, indirectly promoting continued Aβ accumulation.

While the synergy of Aβ-Tau-Inflammation drives disease progression, the catastrophic failure of synaptic integrity and plasticity—the direct structural correlate of cognitive decline—unfolds through a coordinated, multi-stage assault. This process constitutes a self-reinforcing synaptic failure loop. The initial insult involves structural and functional sabotage:[115] Aβ oligomers directly bind to and corrupt postsynaptic densities, while pathological tau disrupts axonal transport, starving synapses of essential components. These initial injuries are dramatically amplified by secondary excitotoxicity and active elimination. The corrupted synaptic environment leads to pathological NMDA receptor overactivation (Section 3.4.2), while the inflamed milieu activates the complement cascade, tagging compromised synapses for microglia-mediated phagocytosis.[116] This aberrant pruning erases neural connections, directly underlying the patient's progressive cognitive decline.

This integrated network model provides a compelling explanation for clinical observations and therapeutic failures. It clarifies why targeting Aβ alone in symptomatic patients fails: by that stage, the tau and neuroinflammation pathways have become self-sustaining drivers of neurodegeneration. Similarly, broadly suppressing inflammation may blunt protective functions, and intervening against tau after its widespread dissemination is likely "too late".

Therefore, the therapeutic imperative is unequivocal: successful disease modification will require multi-target, combination strategies that are initiated early, based on biomarker profiles.[17,117] The model dictates that the optimal window for intervention is before the system reaches criticality—that is, during the preclinical or prodromal stages when Aβ pathology is present, but tau pathology and inflammation are not yet widespread and self-sustaining.[118]

This network perspective directly informs the next generation of clinical trials and the diagnostic paradigm shift discussed in Chapter 4. For example, a rational combination regimen might include: an anti-Aβ monoclonal antibody (e.g., lecanemab) to reduce the initial driver; a tau aggregation inhibitor to halt the spread of tangles; and a precision immunomodulator (e.g., a TREM2 agonist) to enhance protective microglial functions. The future of AD management lies in deconstructing this pathogenic network through preemptive, biomarker-guided application of such personalized, multi-pronged therapeutic assaults.

## 3. Diagnose

Traditional diagnostic paradigms for Alzheimer's disease (AD), such as the NIA-AA criteria,

primarily rely on clinical symptoms and single biomarkers (e.g., Aβ-PET), which are insufficient for capturing the networked nature of the disease and its clinical heterogeneity. Therefore, this paper proposes a new diagnostic framework aimed at guiding diagnosis through an integrated multidimensional network model. We will shift from a dichotomous "presence/absence" framework for single biomarkers to a quantitative and integrated assessment of three core pathological subsystems (Aβ, Tau, neuroinflammation), focusing on their activity, interaction strength, and network-wide effects.

**3.1. Aβ**

In the traditional diagnostic framework (such as AT(N)), the assessment of Aβ typically follows a "binary qualitative" approach. Specifically, through Aβ-PET or cerebrospinal fluid tests, subjects are simply categorized as "Aβ-positive" or "Aβ-negative," primarily based on the overall SUVr value from PET imaging or the total concentration of Aβ42 in the cerebrospinal fluid. Under this diagnostic model, a higher Aβ load is generally interpreted as indicating a greater disease risk or severity. This approach implicitly operates on the assumption that once Aβ pathology exceeds a certain threshold, its pathogenic risk becomes homogeneous, thereby overlooking the specific nodal roles that different brain regions play within neural networks. A growing body of clinical evidence suggests that the quantity of Aβ is far less important than its quality and location. More specifically, the spatial distribution of Aβ and the ratio of soluble oligomers to insoluble plaques are pivotal determinants of whether and how it drives the propagation of Tau pathology.

Under conventional diagnostic frameworks such as AT(N), the assessment of Aβ has generally followed a "binary qualitative" approach. That is, through Aβ-PET or cerebrospinal fluid (CSF) testing, individuals are simply categorized as either "Aβ-positive" or "Aβ-negative," based primarily on the global SUVr value from PET imaging or the overall concentration of Aβ42 in the CSF. Within this diagnostic paradigm, a higher Aβ load is typically interpreted as indicating greater disease risk or severity. This model implicitly assumes that once Aβ pathology exceeds a certain threshold, its associated pathogenic risk is homogeneous—overlooking the specific nodal roles that different brain regions play within neural networks. A growing body of clinical evidence suggests that the quantity of Aβ is far less critical than its quality and anatomical distribution. Specifically, both the spatial distribution of Aβ and the ratio of soluble oligomers to insoluble plaques are pivotal factors in determining whether and how Aβ can drive the spread of Tau pathology.

The spatial distribution of Aβ—particularly its infiltration into specific, highly interconnected neural network hubs—governs the rate at which it drives Tau pathology. When Aβ accumulates in neocortical and limbic regions that maintain strong connectivity with the entorhinal cortex, it can interact remotely with Tau present within the entorhinal cortex. This transregional engagement induces biophysical alterations in entorhinal Tau, enhancing its propensity for aggregation and propagation. As a result, Tau gains the ability to breach its original anatomical confines—the entorhinal cortex—and spread to adjacent memory-critical regions such as the hippocampus and amygdala. Therefore, detecting Aβ within these connected regions of the entorhinal network serves as a more precise early warning sign of impending accelerated Tau spread than relying on global Aβ burden alone. The convergence of Aβ and Tau pathology in the inferior temporal gyrus (ITG) represents a critical juncture for disease acceleration. The ITG is a highly interconnected cortical hub. When Tau pathology, following Braak staging, reaches the ITG and encounters pre-existing Aβ deposits, it creates a potent local inflammatory and toxic microenvironment. This interaction markedly accelerates Tau aggregation and facilitates its subsequent widespread dissemination into neocortical regions. Consequently, the co-localization of Aβ and Tau within hub regions such as the ITG provides direct evidence that the disease has entered a "synergistic acceleration phase."

Conventional diagnostic approaches using Aβ-PET tracers primarily assess the overall burden of insoluble Aβ plaques, a focus historically derived from the therapeutic objective of plaque clearance. However, accumulating evidence indicates that soluble Aβ oligomers—rather than insoluble fibrillar deposits—serve as the principal toxic species driving Tau pathology and neurotoxicity. Insoluble plaques may instead represent a downstream outcome of the disease process or even function as reservoirs that temporarily sequester oligomeric forms. As previously discussed, soluble Aβ oligomers (as opposed to insoluble fibrils) can specifically bind to neuronal surface receptors such as the NMDA receptor and insulin receptor, thereby activating downstream kinase cascades (e.g., GSK-3β, CDK5) that directly induce abnormal hyperphosphorylation of Tau. This mechanism represents the most direct molecular link through which Aβ pathology propels Tau pathogenesis. Thus, we hypothesize that the progression of Alzheimer's disease pathology may depend on the dynamic equilibrium among Aβ monomers, oligomers, and fibrils/plaques. A state characterized by a high oligomer-to-plaque ratio suggests active, ongoing generation of toxic species that continuously assault neurons and activate glial cells. Conversely, a low oligomer-to-plaque ratio may indicate a stabilization of the pathological process, potentially due to effective phagocytic clearance by microglia. This model explains the limited clinical efficacy of plaque-clearing therapies—they fail to effectively eliminate the continuously produced, toxic soluble oligomers. In contrast, newer-generation therapeutics such as Lecanemab and Donanemab represent a significant advance, as they have been shown to more effectively target soluble Aβ aggregates, including protofibrils and oligomers, rather than merely clearing insoluble plaques. This refined targeting likely underlies the more consistent clinical benefits observed in recent trials.

### 3.2. Tau

Traditionally, the Braak staging system has been regarded as a tool for describing the cumulative and unidirectional spread of tau pathology. A patient's Braak stage can be used to indicate disease severity and support diagnosis. However, this staging system has two major limitations: first, it does not adequately explain why tau spreads to specific regions or the rate at which it affects additional areas; second, it treats tau pathology as an isolated process, failing to fully account for how the presence of Aβ fundamentally alters tau propagation and overlooks critical comorbid factors. For example, a Braak stage III patient with substantial Aβ deposits in the temporal regions is likely to experience significantly faster disease progression than a Braak stage III patient without co-existing Aβ pathology. Yet, the Braak staging system cannot capture this distinction. Therefore, this article proposes a reinterpretation of Braak staging results, aiming to provide a basis for precise intervention in the disease course. Braak staging will be regarded as a benchmark for assessing the dynamics of pathological networks and predicting future risks.

Traditionally, Braak stages I-II are considered to represent mild pathological changes in Alzheimer's disease. However, this study focuses on "the presence of Aβ within the entorhinal cortex-connected regions" as a key predictor of the patient's disease trajectory. If Aβ is present in these entorhinal-connected areas, it signifies that Aβ has already begun to drive the outward spread of tau, and the disease course is poised to enter an accelerated phase. Conversely, if Aβ is absent from the entorhinal cortex-connected regions, the disease is likely to progress slowly. In the assessment process, novel MRI techniques, such as dynamic contrast-enhanced MRI, can be employed to evaluate the integrity of the blood-brain barrier (BBB) and investigate the interaction between Aβ and the entorhinal cortex. Research has confirmed that BBB dysfunction is an early event in AD, occurring in specific brain regions. The presence of focal BBB leakage in the entorhinal cortex, coupled with an increased global Aβ load, suggests that a dysfunctional BBB may be failing to clear Aβ effectively, thereby exacerbating its local accumulation. This local aggregation of Aβ creates a toxic microenvironment, which in turn provides the conditions that drive the propagation of tau pathology.

### 3.3. Neuroinflammation

Microglia are capable of phagocytosing Aβ plaques, with the TREM2 protein playing a key regulatory role in this process: under physiological or early pathological conditions, it promotes the clearance of tau protein; however, in later stages, aberrant TREM2 signaling may paradoxically exacerbate tau propagation, thereby posing challenges for therapeutic strategies. Although substantial evidence links neuroinflammation to Alzheimer's disease (AD) pathology, limited longitudinal data and a scarcity of very early biomarkers have hindered the precise determination of causal directionality and disease onset timing. Based on the model proposed in this study—which emphasizes the role of neuroinflammation in accelerating disease progression—we further incorporate neuroinflammation into the diagnostic framework. Through advanced detection techniques, it is now feasible to perform functional subtyping of neuroinflammatory activity, thereby enabling more accurate disease staging and targeted intervention.

As noted above, enhanced TREM2 signaling is associated with Aβ clearance, restricted plaque-associated neurodegeneration, and the maintenance of metabolic homeostasis. Driven by TREM2, microglia migrate to the vicinity of plaques, encapsulating and phagocytosing them, thereby protecting the brain from further plaque-induced damage. Consequently, the detection of a protective inflammatory state in diagnostics indicates that the brain's self-repair mechanisms remain actively engaged. This may correspond to an early disease stage or identify a patient subtype with slower disease progression and stronger cognitive reserve, for whom TREM2 agonists could be therapeutically beneficial. In contrast, destructive neuroinflammation represents a chronic, dysregulated, and intrinsically toxic immune response, characterized fundamentally by the disruption of normal neural function. Danger signals such as Aβ and tau aggregates can activate the NLRP3 inflammasome, leading to caspase-1 activation and the robust release of potent pro-inflammatory cytokines like IL-1β and IL-18. Sustained elevated levels of TNF-α, IL-1β, and IL-6 not only directly harm neurons but also activate kinases such as GSK-3β, directly exacerbating the hyperphosphorylation of tau. Finally, persistent activation transforms microglia, impairing their phagocytic clearance function and converting them into primarily cytokine-secreting "saboteurs" that contribute to aberrant complement-mediated synaptic pruning. Therefore, detecting a destructive inflammatory state signifies accelerated disease progression and may represent a primary reason for rapid cognitive decline even in the presence of low Aβ levels. In this context, NLRP3 inhibitors are warranted to ameliorate the pathology.

As emphasized, determining the state of neuroinflammation is critical for defining the dynamic progression of Alzheimer's disease. While first-generation probes such as TSPO-PET are unable to subtype inflammatory states, a new generation of specific probes shows promise in achieving this. TREM2-PET enables direct imaging of the TREM2 receptor on microglia. Regional elevation of TREM2 signal may indicate the initiation of a protective response, whereas its deficiency or dysfunction may suggest impaired clearance capacity and disease advancement. This probe allows for the functional assessment of TREM2's protective role. In parallel, inflammasome-targeted probes (e.g., NLRP3-PET) directly visualize the NLRP3 inflammasome. A positive signal reflects a "destructive" neuroinflammatory state, thereby providing a precise target for NLRP3 inhibitor therapies aimed at mitigating disease progression. Furthermore, ultrasensitive detection technologies such as SIMOA offer the potential to detect low-concentration inflammatory biomarkers at a lower cost, enabling large-scale screening and dynamic monitoring of therapeutic efficacy.

1       Hardy, J. A. & Higgins, G. A. Alzheimer's disease: the amyloid cascade hypothesis. *Science* **256**, 184–185 (1992).
2       Prince, M. *et al. World Alzheimer report 2015. The global impact of dementia: an analysis of prevalence, incidence, cost and trends*, Alzheimer's Disease International, (2015).
3       Selkoe, D. J. & Hardy, J. The amyloid hypothesis of Alzheimer's disease at 25 years. *EMBO molecular medicine* **8**, 595–608 (2016).
4       Ayton, S. & Bush, A. I. β-amyloid: The known unknowns. *Ageing Research Reviews* **65**, 101212 (2021).
5       Andrade-Moraes, C. H. *et al.* Cell number changes in Alzheimer's disease relate to dementia, not to plaques and tangles. *Brain* **136**, 3738–3752 (2013).
6       Ayton, S. *et al.* Brain iron is associated with accelerated cognitive decline in people with Alzheimer pathology. *Molecular psychiatry* **25**, 2932–2941 (2020).
7       Greene, J. A. & Loscalzo, J.    Vol. 377     2493–2499 (Mass Medical Soc, 2017).
8       Kepp, K. P., Robakis, N. K., Høilund-Carlsen, P. F., Sensi, S. L. & Vissel, B. The amyloid cascade hypothesis: an updated critical review. *Brain* **146**, 3969–3990 (2023).
9       Zhang, J. *et al.* Recent advances in Alzheimer's disease: Mechanisms, clinical trials and new drug development strategies. *Signal transduction and targeted therapy* **9**, 211 (2024).
10      Organization, W. H. *A blueprint for dementia research*.    (World Health Organization, 2022).
11      Hodson, R. Alzheimer's disease. *Nature* **559**, S1, doi:10.1038/d41586-018-05717-6 (2018).
12      Selkoe, D. J. Biochemistry and molecular biology of amyloid beta-protein and the mechanism of Alzheimer's disease. *Handb Clin Neurol* **89**, 245–260, doi:10.1016/s0072-9752(07)01223-7 (2008).
13      Busche, M. A. & Hyman, B. T. Synergy between amyloid-β and tau in Alzheimer's disease. *Nat Neurosci* **23**, 1183–1193, doi:10.1038/s41593-020-0687-6 (2020).
14      Gandy, S. & DeKosky, S. T. Toward the treatment and prevention of Alzheimer's disease: rational strategies and recent progress. *Annual review of medicine* **64**, 367–383 (2013).
15      Hung, S.-Y. & Fu, W.-M. Drug candidates in clinical trials for Alzheimer's disease. *Journal of biomedical science* **24**, 47 (2017).
16      Cummings, J., Lee, G., Ritter, A. & Zhong, K. Alzheimer's disease drug development pipeline: 2018. *Alzheimer's & Dementia: Translational Research & Clinical Interventions* **4**, 195–214 (2018).
17      Cummings, J., Lee, G., Ritter, A., Sabbagh, M. & Zhong, K. Alzheimer's disease drug development pipeline: 2019. *Alzheimer's & Dementia: Translational Research & Clinical Interventions* **5**, 272–293 (2019).

18  Panza, F., Lozupone, M., Logroscino, G. & Imbimbo, B. P. A critical appraisal of amyloid-β-targeting therapies for Alzheimer disease. *Nature Reviews Neurology* **15**, 73–88 (2019).

19  Small, S. A. & Duff, K. Linking Aβ and tau in late-onset Alzheimer's disease: a dual pathway hypothesis. *Neuron* **60**, 534–542 (2008).

20  Iqbal, K., Liu, F., Gong, C.-X. & Grundke-Iqbal, I. Tau in Alzheimer disease and related tauopathies. *Current Alzheimer Research* **7**, 656–664 (2010).

21  Wang, Y. & Mandelkow, E. Tau in physiology and pathology. *Nature reviews neuroscience* **17**, 22–35 (2016).

22  KoSIK, K. S., Joachim, C. L. & Selkoe, D. J. Microtubule-associated protein tau (tau) is a major antigenic component of paired helical filaments in Alzheimer disease. *Proceedings of the National Academy of Sciences* **83**, 4044–4048 (1986).

23  Ittner, L. M. & Götz, J. Amyloid-β and tau—a toxic pas de deux in Alzheimer's disease. *Nature Reviews Neuroscience* **12**, 67–72 (2011).

24  Makin, S. The amyloid hypothesis on trial. *Nature* **559**, S4–S4 (2018).

25  Giannakopoulos, P. *et al.* Tangle and neuron numbers, but not amyloid load, predict cognitive status in Alzheimer's disease. *Neurology* **60**, 1495–1500 (2003).

26  Jack Jr, C. R. *et al.* Update on hypothetical model of Alzheimer's disease biomarkers. *Lancet neurology* **12**, 207 (2013).

27  Bennett, D. *et al.* Neuropathology of older persons without cognitive impairment from two community-based studies. *Neurology* **66**, 1837–1844 (2006).

28  Katzman, R. *et al.* Clinical, pathological, and neurochemical changes in dementia: a subgroup with preserved mental status and numerous neocortical plaques. *Annals of Neurology: Official Journal of the American Neurological Association and the Child Neurology Society* **23**, 138–144 (1988).

29  Serrano-Pozo, A., Frosch, M. P., Masliah, E. & Hyman, B. T. Neuropathological alterations in Alzheimer disease. *Cold Spring Harbor perspectives in medicine* **1**, a006189 (2011).

30  Vasconcelos, B. *et al.* Heterotypic seeding of Tau fibrillization by pre-aggregated Abeta provides potent seeds for prion-like seeding and propagation of Tau-pathology in vivo. *Acta neuropathologica* **131**, 549–569 (2016).

31  Avila, J., Hernández, F. & Perry, G. Amyloid and tau pathologies cross-talk to promote Alzheimeŕs disease: novel mechanistic insights. *Neuroscience* (2025).

32  He, Z. *et al.* Amyloid-β plaques enhance Alzheimer's brain tau-seeded pathologies by facilitating neuritic plaque tau aggregation. *Nature Medicine* **24**, 29–38, doi:10.1038/nm.4443 (2018).

33  Pooler, A. M. *et al.* Amyloid accelerates tau propagation and toxicity in a model of early Alzheimer's disease. *Acta Neuropathol Commun* **3**, 14, doi:10.1186/s40478-015-0199-x (2015).

34  Bloom, G. S. Amyloid-β and tau: the trigger and bullet in Alzheimer disease pathogenesis. *JAMA Neurol* **71**, 505–508, doi:10.1001/jamaneurol.2013.5847 (2014).

35  McDade, E. *et al.* Longitudinal cognitive and biomarker changes in dominantly inherited Alzheimer disease. *Neurology* **91**, e1295–e1306, doi:doi:10.1212/WNL.0000000000006277 (2018).

36  Jack, C. R., Jr. *et al.* Hypothetical model of dynamic biomarkers of the Alzheimer's pathological cascade. *The Lancet Neurology* **9**, 119–128, doi:10.1016/S1474-4422(09)70299-6 (2010).

37  Lee, W. J. *et al.* Regional Aβ-tau interactions promote onset and acceleration of Alzheimer's disease tau spreading. *Neuron* **110**, 1932–1943.e1935, doi:10.1016/j.neuron.2022.03.034 (2022).

38  郑晴晴 & 刘煜. Tau 蛋白在阿尔茨海默病中的发病机制及治疗策略进展. *Pharmacy Information* **14**, 129 (2025).

39  Zheng, W.-H., Bastianetto, S., Mennicken, F., Ma, W. & Kar, S. Amyloid β peptide induces tau phosphorylation and loss of cholinergic neurons in rat primary septal cultures. *Neuroscience* **115**, 201–211 (2002).

40  Iijima, K., Gatt, A. & Iijima-Ando, K. Tau Ser262 phosphorylation is critical for Aβ42-induced tau toxicity in a transgenic Drosophila model of Alzheimer's disease. *Human molecular genetics* **19**, 2947–2957 (2010).

41  Grundke-Iqbal, I. *et al.* Abnormal phosphorylation of the microtubule-associated protein tau (tau) in Alzheimer cytoskeletal pathology. *Proceedings of the National Academy of Sciences* **83**, 4913–4917 (1986).

42  Zhang, H. *et al.* Interaction between Aβ and tau in the pathogenesis of Alzheimer's disease. *International journal of biological sciences* **17**, 2181 (2021).

43  Ryan, S. D. *et al.* Amyloid-β42 signals tau hyperphosphorylation and compromises neuronal viability by disrupting alkylacylglycerophosphocholine metabolism. *Proceedings of the National Academy of Sciences* **106**, 20936–20941 (2009).

44  Song, J.-S. & Yang, S.-D. Tau protein kinase I/GSK-3β/kinase FA in heparin phosphorylates tau on Ser199, Thr231, Ser235, Ser262, Ser369, and Ser400 sites phosphorylated in Alzheimer disease brain. *Journal of protein chemistry* **14**, 95–105 (1995).

45  Feng, Y. *et al.* Cleavage of GSK-3β by calpain counteracts the inhibitory effect of Ser9 phosphorylation on GSK-3β activity induced by H2O2. *Journal of neurochemistry* **126**, 234–242 (2013).

46  Hromadkova, L. *et al.* Structural exposure of different microtubule binding domains determines the propagation and toxicity of pathogenic tau conformers in Alzheimer's disease. *PLoS pathogens* **21**, e1012926 (2025).

47  Gerson, J. E. & Kayed, R. Formation and propagation of tau oligomeric seeds. *Frontiers in neurology* **4**, 93 (2013).

48  Takashima, A. GSK-3 is essential in the pathogenesis of Alzheimer's disease. *Journal of Alzheimer's disease* **9**, 309–317 (2006).

49  Nwadiugwu, M., Onwuekwe, I., Ezeanolue, E. & Deng, H. Beyond amyloid: a machine learning-driven approach reveals properties of potent GSK-3β inhibitors targeting neurofibrillary tangles. *International Journal of Molecular Sciences* **25**, 2646 (2024).

50  Mintun, M. A. *et al.* Donanemab in early Alzheimer's disease. *New England Journal of Medicine* **384**, 1691–1704 (2021).

51  Manczak, M. & Reddy, P. H. Abnormal interaction of oligomeric amyloid-β with phosphorylated tau: implications to synaptic dysfunction and neuronal damage. *Journal of Alzheimer's Disease* **36**, 285–295 (2013).

52	Evans, C. E. *et al.* Selective reduction of APP-BACE1 activity improves memory via NMDA-NR2B receptor-mediated mechanisms in aged PDAPP mice. *Neurobiology of Aging* **75**, 136–149 (2019).

53	Wang, B. *Amyloid Beta Peptide Induces D-serine Dependent NMDAR Dysfunction in the Mouse Hippocampus*, Université d'Ottawa/University of Ottawa, (2016).

54	Koffie, R. M., Hyman, B. T. & Spires-Jones, T. L. Alzheimer's disease: synapses gone cold. *Molecular neurodegeneration* **6**, 63 (2011).

55	Zhang, Z. *et al.* Blocking the NR2B in the hippocampal dentate gyrus reduced the spatial memory deficits and apoptosis through the PERK-CHOP pathway in a rat model of sporadic Alzheimer's disease. *Behavioural Brain Research*, 115685 (2025).

56	Roberson, E. D. *et al.* Reducing endogenous tau ameliorates amyloid β-induced deficits in an Alzheimer's disease mouse model. *Science* **316**, 750–754 (2007).

57	Amin, J. *et al.* Effect of amyloid-β (A β) immunization on hyperphosphorylated tau: a potential role for glycogen synthase kinase (GSK)-3β. *Neuropathology and applied neurobiology* **41**, 445–457 (2015).

58	Roberson, E. D. *et al.* Amyloid-β/Fyn–induced synaptic, network, and cognitive impairments depend on tau levels in multiple mouse models of Alzheimer's disease. *Journal of Neuroscience* **31**, 700–711 (2011).

59	Stamer, K., Vogel, R., Thies, E., Mandelkow, E. & Mandelkow, E.-M. Tau blocks traffic of organelles, neurofilaments, and APP vesicles in neurons and enhances oxidative stress. *The Journal of cell biology* **156**, 1051–1063 (2002).

60	Mandelkow, E.-M., Stamer, K., Vogel, R., Thies, E. & Mandelkow, E. Clogging of axons by tau, inhibition of axonal traffic and starvation of synapses. *Neurobiology of aging* **24**, 1079–1085 (2003).

61	Zhang, Z. *et al.* δ-Secretase-cleaved Tau stimulates Aβ production via upregulating STAT1-BACE1 signaling in Alzheimer's disease. *Molecular psychiatry* **26**, 586–603 (2021).

62	Luan, Y. *et al.* Synaptic density and tau pathology in Alzheimer's disease: a dual role in susceptibility and degeneration. *Brain*, doi:10.1093/brain/awaf425 (2025).

63	Busche, M. A. & Hyman, B. T. Synergy between amyloid-β and tau in Alzheimer's disease. *Nature neuroscience* **23**, 1183–1193 (2020).

64	Busche, M. A. *et al.* Tau impairs neural circuits, dominating amyloid-β effects, in Alzheimer models in vivo. *Nature neuroscience* **22**, 57–64 (2019).

65	Wang, X., Ye, T., Zhou, W., Zhang, J. & Initiative, A. s. D. N. Association of Distinct Initial β-Amyloid Levels With Tau Pathology Expansion Beyond the Entorhinal Cortex. *Neurology* **105**, e214041 (2025).

66	Pascoal, T. A. *et al.* Amyloid-β and hyperphosphorylated tau synergy drives metabolic decline in preclinical Alzheimer's disease. *Molecular psychiatry* **22**, 306–311 (2017).

67	Gallego-Rudolf, J. *et al.* Synergistic association of Aβ and tau pathology with cortical neurophysiology and cognitive decline in asymptomatic older adults. *Nature Neuroscience* **27**, 2130–2137 (2024).

68	Lewis, J. *et al.* Enhanced neurofibrillary degeneration in transgenic mice expressing mutant tau and APP. *Science* **293**, 1487–1491 (2001).

69  Gotz, J., Chen, F. v., Van Dorpe, J. & Nitsch, R. Formation of neurofibrillary tangles in P301L tau transgenic mice induced by Aβ42 fibrils. *Science* **293**, 1491–1495 (2001).

70  He, Z. *et al.* Amyloid-β plaques enhance Alzheimer's brain tau-seeded pathologies by facilitating neuritic plaque tau aggregation. *Nature medicine* **24**, 29–38 (2018).

71  Han, Y.-E., Lim, S., Lee, S. E., Nam, M.-H. & Oh, S.-J. Novel mouse model of Alzheimer's disease exhibits pathology through synergistic interactions among amyloid-β, tau, and reactive astrogliosis. *Zoological Research* **46**, 41 (2025).

72  Bennett, R. E. *et al.* Enhanced tau aggregation in the presence of amyloid β. *The American journal of pathology* **187**, 1601–1612 (2017).

73  Leng, F. & Edison, P. Neuroinflammation and microglial activation in Alzheimer disease: where do we go from here? *Nature Reviews Neurology* **17**, 157–172 (2021).

74  Heppner, F. L., Ransohoff, R. M. & Becher, B. Immune attack: the role of inflammation in Alzheimer disease. *Nature Reviews Neuroscience* **16**, 358–372 (2015).

75  Heneka, M. T., Kummer, M. P. & Latz, E. Innate immune activation in neurodegenerative disease. *Nature Reviews Immunology* **14**, 463–477 (2014).

76  Hickman, S. E., Allison, E. K. & El Khoury, J. Microglial dysfunction and defective β-amyloid clearance pathways in aging Alzheimer's disease mice. *Journal of Neuroscience* **28**, 8354–8360 (2008).

77  Halle, A. *et al.* The NALP3 inflammasome is involved in the innate immune response to amyloid-β. *Nature immunology* **9**, 857–865 (2008).

78  Venegas, C. *et al.* Microglia-derived ASC specks cross-seed amyloid-β in Alzheimer's disease. *Nature* **552**, 355–361 (2017).

79  Ghosh, S. *et al.* Sustained interleukin-1β overexpression exacerbates tau pathology despite reduced amyloid burden in an Alzheimer's mouse model. *Journal of Neuroscience* **33**, 5053–5064 (2013).

80  Hanger, D. P., Anderton, B. H. & Noble, W. Tau phosphorylation: the therapeutic challenge for neurodegenerative disease. *Trends in molecular medicine* **15**, 112–119 (2009).

81  Heneka, M. T. *et al.* Neuroinflammation in Alzheimer's disease. *The Lancet Neurology* **14**, 388–405 (2015).

82  De Strooper, B. & Karran, E. The cellular phase of Alzheimer's disease. *Cell* **164**, 603–615 (2016).

83  Wang, E. *et al.* Multiscale proteomic modeling reveals protein networks driving Alzheimer's disease pathogenesis. *Cell* (2025).

84  Mrdjen, D. *et al.* Spatial proteomics of Alzheimer's disease-specific human microglial states. *Nature Immunology*, 1–14 (2025).

85  Zheng, X. *et al.* Neural Stem Cell-conditioned Medium Protected Against Cognitive Dysfunction in APP/PS1 Mice. *Mol Neurobiol* **63**, 27, doi:10.1007/s12035-025-05282-w (2025).

86  Shademan, B., Yousefi, H., Sharafkhani, R. & Nourazarian, A. LPS-Induced Neuroinflammation Disrupts Brain-Derived Neurotrophic Factor and Kinase Pathways in Alzheimer's Disease Cell Models. *Cell Mol Neurobiol* **45**, 102, doi:10.1007/s10571-025-01600-x (2025).

87      Ayata, P. *et al.* Lymphoid gene expression supports neuroprotective microglia function. *Nature*, doi:10.1038/s41586-025-09662-z (2025).

88      Johnson, A. M. & Lukens, J. R. Emerging roles for innate and adaptive immunity in tauopathies. *Cell Rep* **44**, 116232, doi:10.1016/j.celrep.2025.116232 (2025).

89      Jha, D., Bakker, E. N. & Kumar, R. Mechanistic and therapeutic role of NLRP3 inflammasome in the pathogenesis of Alzheimer's disease. *Journal of Neurochemistry* **168**, 3574–3598 (2024).

90      Shang, H., Duan, M., Da, C., Fu, X. & Zhang, T. Cross-talk between amyloid beta peptides and tau proteins in Co-aggregation investigating with the combination of coarse-grained and all-atom simulations. *International Journal of Biological Macromolecules*, 144651 (2025).

91      Rajmohan, R. & Reddy, P. H. Amyloid-beta and phosphorylated tau accumulations cause abnormalities at synapses of Alzheimer's disease neurons. *Journal of Alzheimer's Disease* **57**, 975–999 (2017).

92      Sun, Z.-D., Hu, J.-X., Wu, J.-R., Zhou, B. & Huang, Y.-P. Toxicities of amyloid-beta and tau protein are reciprocally enhanced in the Drosophila model. *Neural regeneration research* **17**, 2286–2292 (2022).

93      Kim, M. *et al.* Interactions with tau's microtubule-binding repeats modulate amyloid-β aggregation and toxicity. *Nature Chemical Biology*, 1–10 (2025).

94      Singh, D. Astrocytic and microglial cells as the modulators of neuroinflammation in Alzheimer's disease. *Journal of neuroinflammation* **19**, 206 (2022).

95      Johnstone, M., Gearing, A. J. & Miller, K. M. A central role for astrocytes in the inflammatory response to β-amyloid; chemokines, cytokines and reactive oxygen species are produced. *Journal of neuroimmunology* **93**, 182–193 (1999).

96      Zhang, L. *et al.* The Mechanism of TNF-α-Mediated Accumulation of Phosphorylated Tau Protein and Its Modulation by Propofol in Primary Mouse Hippocampal Neurons: Role of Mitophagy, NLRP3, and p62/Keap1/Nrf2 Pathway. *Oxidative Medicine and Cellular Longevity* **2022**, 8661200 (2022).

97      Sen, T., Saha, P., Jiang, T. & Sen, N. Sulfhydration of AKT triggers Tau-phosphorylation by activating glycogen synthase kinase 3β in Alzheimer's disease. *Proceedings of the National Academy of Sciences* **117**, 4418–4427 (2020).

98      Sheng, J. G. *et al.* Interleukin-1 promotion of MAPK-p38 overexpression in experimental animals and in Alzheimer's disease: potential significance for tau protein phosphorylation. *Neurochemistry international* **39**, 341–348 (2001).

99      Mammeri, N. E., Dregni, A. J., Duan, P. & Hong, M. Structures of AT8 and PHF1 phosphomimetic tau: Insights into the posttranslational modification code of tau aggregation. *Proceedings of the National Academy of Sciences* **121**, e2316175121 (2024).

100     张梦阳 & 王玥. 糖原合酶激酶-3β 在阿尔茨海默病发病机制及治疗中作用的研究概况. *中华诊断学电子杂志* **7**, 98 (2019).

101     Li, Y., Liu, L., Barger, S. W. & Griffin, W. S. T. Interleukin-1 mediates pathological effects of microglia on tau phosphorylation and on synaptophysin synthesis in cortical neurons through a p38-MAPK pathway. *Journal of Neuroscience* **23**, 1605–1611 (2003).

102 Leyns, C. E. *et al.* TREM2 function impedes tau seeding in neuritic plaques. *Nature neuroscience* **22**, 1217–1222 (2019).

103 Guo, F. *et al.* Complement C1q is associated with neuroinflammation and mediates the association between amyloid-β and tau pathology in Alzheimer's disease. *Translational Psychiatry* **15**, 247 (2025).

104 Han, Q.-Q., Shen, S.-Y., Liang, L.-F., Chen, X.-R. & Yu, J. Complement C1q/C3-CR3 signaling pathway mediates abnormal microglial phagocytosis of synapses in a mouse model of depression. *Brain, Behavior, and Immunity* **119**, 454–464 (2024).

105 Ayyubova, G. & Fazal, N. Beneficial versus Detrimental Effects of Complement–Microglial Interactions in Alzheimer's Disease. *Brain Sciences* **14**, 434 (2024).

106 Schafer, D. P. *et al.* Microglia sculpt postnatal neural circuits in an activity and complement-dependent manner. *Neuron* **74**, 691–705 (2012).

107 Nybo, M., Nielsen, E. H. & Svehag, S.-E. Complement activation by the amyloid proteins Aβ peptide and β2-microglobulin. *Amyloid* **6**, 265–272 (1999).

108 Rogers, J. *et al.* Complement activation by beta-amyloid in Alzheimer disease. *Proceedings of the National Academy of Sciences* **89**, 10016–10020 (1992).

109 Ng, A. N., Salter, E. W., Georgiou, J., Bortolotto, Z. A. & Collingridge, G. L. Amyloid-β1-42 oligomers enhance mGlu5R-dependent synaptic weakening via NMDAR activation and complement C5aR1 signaling. *Iscience* **26** (2023).

110 Doig, A. J. Positive feedback loops in Alzheimer's disease: the Alzheimer's feedback hypothesis. *Journal of Alzheimer's Disease* **66**, 25–36 (2018).

111 Sanchez-Rodriguez, L. M. *et al.* Personalized whole-brain neural mass models reveal combined Aβ and tau hyperexcitable influences in Alzheimer's disease. *Communications Biology* **7**, 528 (2024).

112 Kiselica, A. M. & Initiative, A. s. D. N. Empirically defining the preclinical stages of the Alzheimer's continuum in the Alzheimer's Disease Neuroimaging Initiative. *Psychogeriatrics* **21**, 491–502 (2021).

113 姜敏艳, 汪旭 & 郭锡汉. 小胶质细胞 TREM2 调控阿尔茨海默病发生的机制研究进展. *生命科学* **35**, 1484–1497 (2023).

114 Jucker, M. & Walker, L. C. Self-propagation of pathogenic protein aggregates in neurodegenerative diseases. *Nature* **501**, 45–51 (2013).

115 Tang, Y. *et al.* Early and selective impairments in axonal transport kinetics of synaptic cargoes induced by soluble amyloid β-protein oligomers. *Traffic* **13**, 681–693 (2012).

116 Zhang, F.-L., Hong, K., Li, T.-J., Park, H. & Yu, J.-Q. Functionalization of C (sp3)–H bonds using a transient directing group. *Science* **351**, 252–256 (2016).

117 Boongird, C. *et al.* Bridging the gap: Efficacy of combined therapies for cognitive, behavioral, and functional outcomes in Alzheimer's disease - results from a systematic review and network meta-analysis. *J Alzheimers Dis* **108**, 509–521, doi:10.1177/13872877251378354 (2025).

118 Vecchio, I. & Colica, C. New Research on Biomarkers in Alzheimer's Continuum. *Rev Recent Clin Trials*, doi:10.2174/0115748871331138250114052615 (2025).